\documentclass{article}

    \usepackage[final,nonatbib]{neurips_2024_ml4ps}

\usepackage[utf8]{inputenc} 
\usepackage[T1]{fontenc}    
\usepackage{hyperref}       
\usepackage{url}            
\usepackage{booktabs}       
\usepackage{amsfonts}       
\usepackage{nicefrac}       
\usepackage{microtype}      
\usepackage{xcolor}         
\usepackage{cite}
\usepackage{latexsym} 	
\usepackage{amssymb} 
\usepackage{amsbsy}
\usepackage{amsmath}
\usepackage{epsfig}     
\usepackage{graphics,color}
\usepackage{multirow}
\usepackage{hyperref}       

\newcommand{\be}{\begin{equation}}
\newcommand{\ee}{\end{equation}}
\newcommand{\bea}{\begin{eqnarray}}
\newcommand{\eea}{\end{eqnarray}}
\newcommand{\bean}{\begin{eqnarray*}}
\newcommand{\eean}{\end{eqnarray*}}
\newcommand{\bit}{\begin{itemize}}
\newcommand{\eit}{\end{itemize}}

\newcommand{\half}{\frac{1}{2}}
\newcommand{\id}{1\!\!1}

\newcommand{\qqquad}{\qquad\qquad}

\newcommand{\cB}{{\cal B}}
\newcommand{\cL}{{\cal L}}
\newcommand{\cN}{{\cal N}}

\newcommand{\ket}{\rangle}
\newcommand{\bra}{\langle}

\usepackage{chngcntr}

\counterwithin*{equation}{section}
\hypersetup{linkcolor=blue}   

\title{Dyson Brownian motion and random matrix dynamics of weight matrices during learning}

\author{%
Gert Aarts \\
Department of Physics, Swansea University, Swansea SA2 8PP, United Kingdom \\
\texttt{g.aarts@swansea.ac.uk} 
\And
Ouraman Hajizadeh \\
Graz, Austria \\
\texttt{ou86hajizadeh@gmail.com}
\And
Biagio Lucini \\
Department of Mathematics, Swansea University, 
Swansea, SA1 8EN, United Kingdom \\
\texttt{b.lucini@swansea.ac.uk} \\
\And
Chanju Park \\
Department of Physics, Swansea University, Swansea SA2 8PP, United Kingdom\\
\texttt{chanju.b.park@gmail.com} \\
\mbox{}\\
November 20, 2024
}

\begin{document}

\maketitle

\begin{abstract}
  During training, weight matrices in machine learning architectures are updated using stochastic gradient descent or variations thereof. In this contribution we employ concepts of random matrix theory to analyse the resulting stochastic matrix dynamics. We first demonstrate that the dynamics can generically be described using Dyson Brownian motion, leading to e.g.\  eigenvalue repulsion. The level of stochasticity is shown to depend on the ratio of the learning rate and the mini-batch size, explaining the empirically observed linear scaling rule. We verify this linear scaling in the restricted Boltzmann machine.  
  Subsequently we study weight matrix dynamics in transformers (a nano-GPT), following the evolution from a Marchenko-Pastur distribution for eigenvalues at initialisation to a combination with additional structure at the end of learning.   
   
\end{abstract}

\section{Introduction}

The dynamics of weight matrices $W$ during training is a topic of considerable interest, as it may shed light on both the capacity and efficiency of machine learning (ML) architectures to learn and generalise. Since most algorithms involve a level of stochasticity, e.g.\ due to the use of mini-batches and sampling, the appropriate manner to analyse this dynamics is random matrix theory (RMT) \cite{Wigner-1,Wigner-2,Dyson-1,Dyson-2,Dyson-3,Meh2004}. 
In the presence of time evolution, as is the case for learning, one can then employ the framework of Dyson Brownian motion \cite{Dyson-4} to express the dynamics of $X=W^TW$  in terms of a stochastic evolution equation for the eigenvalues of $X$ \cite{Aarts:2024wxi}. This equation includes an induced Coulomb term, resulting in e.g.\ eigenvalue repulsion and the Wigner surmise.

Here we address the question how Dyson Brownian motion can be applied to stochastic weight matrix updates, with observable consequences. Reducing the learning dynamics from the space of matrices to the space of eigenvalues naturally yields the picture in which the eigenvalues evolve from an initial Marchenko-Pastur distribution to one determined by a so-called stochastic Coulomb potential. This Coulomb potential contains both universal and non-universal aspects, which depend on details of the minimisation landscape due to the loss function. Hence a study of the evolution of eigenvalues provides insight into the latter. To demonstrate our analytical findings, we present results in the Gaussian restricted Boltzmann machine (RBM), which can be understood completely, and for transformers (a nano-GPT), where results are currently at an empirical level. 

\paragraph{Related work} RMT dates back to the analysis of nuclear spectra in the 1960s \cite{Wigner-1,Wigner-2,Dyson-1,Dyson-2,Dyson-3,Meh2004}, but has been proven to be widely applicable across disciplines. 
The notion that RMT is useful for the description of weight matrices in ML has been observed previously, see e.g.\ Refs.\  \cite{pennington2017nonlinear,Martin-2019,Baskerville,Baskerville_2022} and the textbook \cite{couillet_liao_2022}. As far as we know the connection between Dyson Brownian motion and the stochastic Coulomb gas on the one hand and the eigenvalue dynamics of weight matrices on the other hand was first pointed in Ref.\ \cite{Aarts:2024wxi}.
RMT has also been used to study properties of data \cite{Levi:2023qwg} instead of weight matrices. 

\section{Dyson Brownian motion and stochastic matrix dynamics}

Stochastic gradient descent, or variations thereof, are widely used to minimise loss functions, $\cL[W]$, via the update
$W \to W' = W+\delta W = W-\alpha \delta \cL/\delta W$, where $\alpha$ is the learning rate. We follow here closely our previous work \cite{Aarts:2024wxi}.
Using stochastically chosen mini-batches $\cB$ of size $|\cB|$ and invoking the central limit theorem, yields the stochastic update for weight matrices $W$ 
\be
\label{eq:Wex}
W_{ij}' = W_{ij} -\alpha\left(\frac{\delta\cal L}{\delta W_{ij}}\right)_{\cal B} +\frac{\alpha}{\sqrt{|{\cal B}|}}
\sqrt{\mbox{Var}\left(\frac{\delta\cal L}{\delta W_{ij}} \right)} \eta_{ij},
\ee
where $\eta_{ij}\sim\cN(0,1)$, the subscript $\mathcal{B}$ denotes the true mean of the batch gradient, and there is no summation over repeated indices. 
To apply RMT, it is beneficial to consider the symmetric combination $X =  W^TW$, which has real and semi-positive eigenvalues (the squares of $W$'s singular values). The update of $X$ can be written as
\be
\label{eq:deltaX}
X_{ij}\to X_{ij}' = X_{ij} +  \delta X^{\cal B}_{ij} + \frac{1}{\sqrt{|{\cal B|}}}\sqrt{\mbox{Var}(\delta X_{ij})} \eta_{ij},
\ee
where the various terms follow from Eq.~(\ref{eq:Wex}) and the noise $\eta_{ij}$ is symmetric in this case.
Note that the actual update is carried out using Eq.~(\ref{eq:Wex}): the semi-positive and symmetric matrix $X$ is constructed at each iteration by choosing $X = W^TW$.
Using now the framework of Dyson Brownian motion \cite{Dyson-4,Meh2004} yields a stochastic equation for the eigenvalues $x_i$ ($i=1,\ldots,N$) of $X$, which reads
\be
\label{eq:x}
x_i\to x_i' = x_i + K_i +\sum_{j\neq i}\frac{g_i^2}{x_i-x_j} +\sqrt{2}g_i\eta_i,
\ee
where $K_i$ and $g_i$ are linked to the deterministic and stochastic terms in Eq.\ (\ref{eq:deltaX}) and again $\eta_i\sim\cN(0,1)$. The Coulomb term results in eigenvalue repulsion. 
Making the learning rate and batch size explicit by writing, c.f.\ Eq.\ (\ref{eq:Wex}), 
$K_i = \alpha \tilde K_i$ and 
$g_i = (\alpha/\sqrt{|{\cal B|}}) \tilde g_i$
 (quantities with a tilde are independent of the learning rate and batch size)
yields the eigenvalue equation
\be
\label{eq:xtilde}
x_i\to x_i' = x_i +  \alpha \tilde K_i + \frac{\alpha^2}{|{\cal B}|}\sum_{j\neq i}\frac{\tilde g_i^2}{x_i-x_j} + 
\frac{\alpha}{\sqrt{|{\cal B}|}}\sqrt{2}\tilde g_i\eta_i.
\ee
At first sight, the scaling with $\alpha$ and $|\cB|$ looks unwieldy. However, it falls into place when considering the stationary distribution corresponding to the stochastic processes (\ref{eq:x}, \ref{eq:xtilde}), which is obtained via the associated Fokker-Planck equation. This so-called Coulomb gas distribution reads
\be
\label{eq:Cgas}
P_s(\{x_i\}) =  \frac{1}{Z}\prod_{i<j} \left|x_i-x_j\right| e^{-\sum_i V_i(x_i)/g_i^2},
\qqquad
Z = \int dx_1 \ldots dx_N\, P_s(\{x_i\}).
\ee
Here it is assumed that the drift $K_i$ can be derived from a separable potential $V_i(x_i)$, via $K_i(x_i) = -dV_i(x_i)/dx_i$.
By writing $V_i(x_i) = \alpha \tilde  V_i(x_i)$, the learning rate and batch size can be made explicit again, and the combination in the exponent  reads
\be
\label{eq:Vi}
\frac{V_i(x_i)}{g_i^2} = \frac{1}{\alpha/|\cB|} \frac{\tilde V_i(x_i)}{\tilde g_i^2}.
\ee
The first factor on the RHS indicates universal scaling with $\alpha/|\cB|$, while the second factor depends on the details of the loss function.
The linear scaling rule, in which $\alpha/|\cB|$ is a parameter whose tunability can be exploited, was previously observed empirically \cite{Smith-1,Smith-2}. Here it is shown that it is a consequence of stochastic matrix dynamics, when cast in the framework of Brownian motion (see Refs.~\cite{DBLP:journals/corr/abs-1710-11029,li2017} for an alternative derivation).
Reducing the level of stochasticity, by reducing $\alpha/|\cB|$, leads to a distribution with spectral weight concentrated more closely around the minima of $V_i(x_i)$.

\section{Applications}

The derivation above is rather general and the actual expressions for the drift $K_i$, the noise strength $g_i$ and the Coulomb potential $V_i$ may not be readily available. In this section, we apply the description to two systems: the Gaussian RBM, in which the potential $V_i(x_i)$ can be given explicitly \cite{Aarts:2024wxi}, and a transformer, in which the analysis is at an empirical level.

\subsection{Gaussian restricted Boltzmann machine}
 
The Gaussian RBM, with continuous degrees of freedom $\phi_i$ ($i=1,\ldots,N_v$) and $h_a$ ($a=1,\ldots,N_h$) on the visible and hidden layer respectively, is determined by the energy (see e.g.\ the review \cite{Decelle_2021}, here we closely follow Refs.~\cite{Aarts:2023uwt,Aarts:2024wxi}) 
\be
E(\phi, h) = \half \mu^2 \phi^T\phi +\frac{1}{2\sigma_h^2}h^Th - \phi^T W h.
\ee
Here $\mu^2$ and $\sigma_h^2$ are hyperparameters and we have put a possible bias to zero. The rectangular weight matrix $W$, with matrix elements $W_{ia}$, connects the layers. 
The induced distribution on the visible layer is determined by the matrix
$K_{\rm RBM} = \mu^2\id_{N_v\times N_v}-\sigma_h^2 WW^T$.
The target data is summarised in an $N_v\times N_v$ matrix with eigenvalues $\kappa_i$.
In the stochastic equation for the eigenvalues $x_i$ of $X=\sigma_h^2 W^TW$, the drift is known explicitly \cite{Aarts:2024wxi},
\be
\label{eq:xRBM}
\frac{d}{d\tau}x_i = K_i(x_i) + \sum_{j\neq i}\frac{g_i^2}{x_i-x_j} + \sqrt{2}g_i\eta_i,
\qqquad
K_i(x_i) = \left( \frac{1}{\kappa_i} -\frac{1}{\mu^2-x_i}\right)x_i,
\ee
where $\tau = 2\sigma_h^2t$ (for notational simplicitly, we use here continuous time).
The parametric dependence of the stochasticity parameter $g_i^2$ can be determined explicitly as well, with
$g_i^2 \sim (\alpha/|\cB|)\times \kappa_i^2\Omega_i$, with $\Omega_i=(\mu^2-\kappa_i)/\kappa_i^2$,
where, as stated above, the first factor is universal and the second factor is model ($\mu^2$) and data ($\kappa_i$) dependent \cite{Aarts:2024wxi}.

\begin{figure}[b]
\begin{center}
\includegraphics[width=0.45\textwidth]{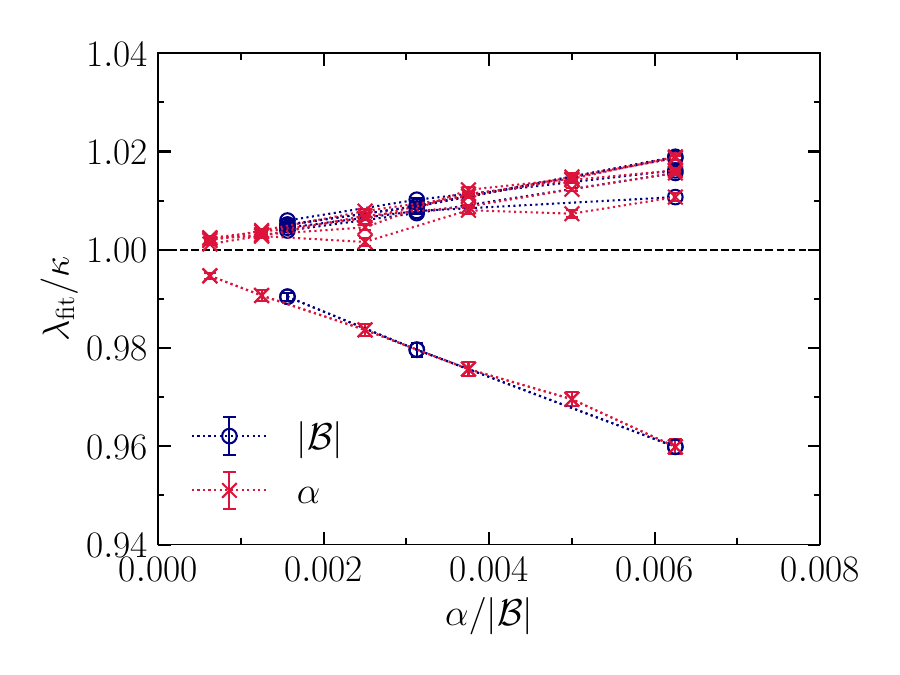}
\includegraphics[width=0.45\textwidth]{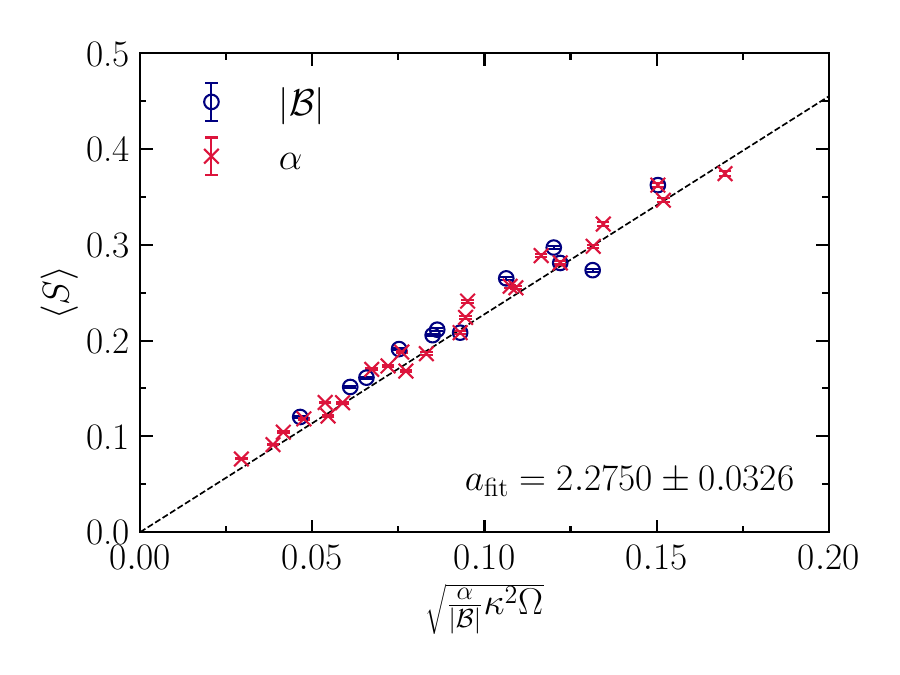}
\end{center}
 \caption{
 Gaussian RBM:
 Ratio of the RBM eigenvalues $\lambda_i=\mu^2-x_i$ and the target eigenvalues $\kappa_i$ as a function of $\alpha/|\cB|$, where $\alpha$ and $|\cB|$ are independently varied,  demonstrating eigenvalue repulsion for non-vanishing stochasticity (left). 
 Response of the mean level spacing $\bra S\ket$ to variation of $\alpha$ and  $|\cB|$, presented in the combination $\sqrt{\alpha/|\cB|}$ times a non-universal function $\sqrt{\kappa_i^2 \Omega_i}=\sqrt{\mu^2-\kappa_i}$ (right).
 Figures from Ref.\ \cite{Aarts:2024wxi}.
 }
\label{fig:RBM}
\end{figure}

The Coulomb term in Eq.\ (\ref{eq:xRBM}) leads to eigenvalue repulsion, with a strength proportional to $\alpha/|\cB|$. 
This is demonstrated in Fig.\ \ref{fig:RBM} (left). Only in the limit where $\alpha/|\cB|\to 0$ will the target spectrum be learnt exactly.
The distribution of the nonzero level spacing $S_i=x_{i+1}-x_i$ is known as the Wigner surmise. The mean level $\bra S\ket$ is proportional to $\sqrt{\alpha/|\cB|}$, as is confirmed in  Fig.\ \ref{fig:RBM} (right). 

\begin{figure}
    \begin{center}
    \includegraphics[width=0.45\linewidth]{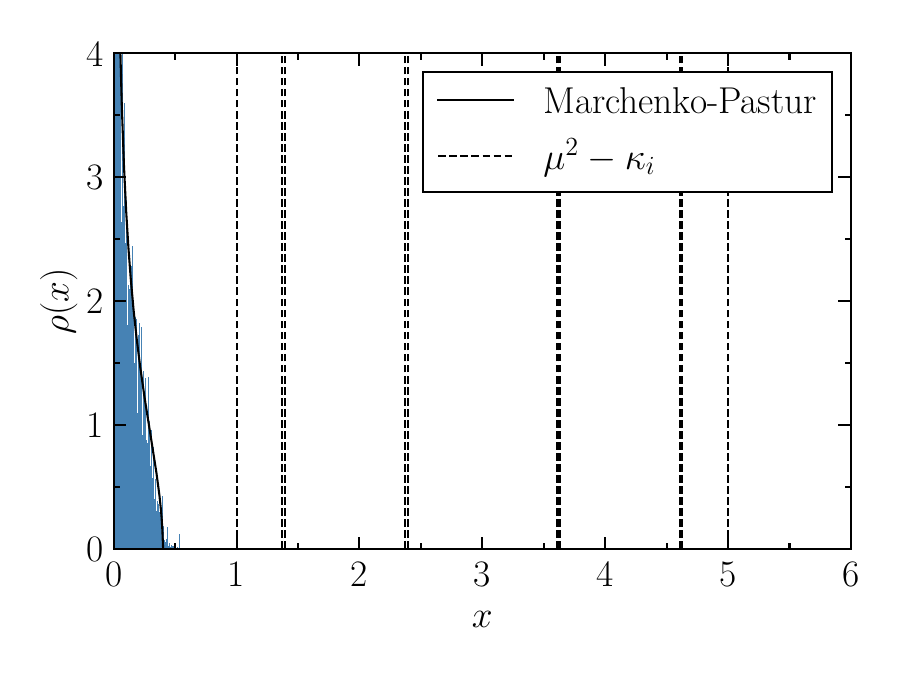}
    \includegraphics[width=0.45\linewidth]{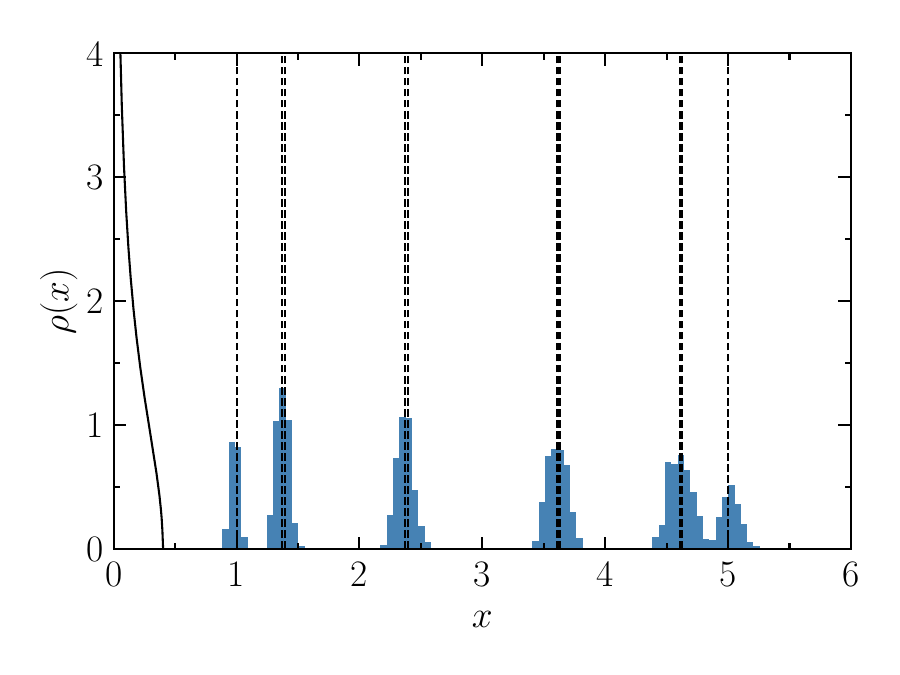}
    \end{center}
    \caption{
    Gaussian RBM: Evolution of eigenvalues of $X = \sigma_h^2 W^TW$ from the Marchenko-Pastur distribution at initialisation (left) to the learned distribution around the target eigenvalues, indicated with the vertical lines, at the end of training (right).
    }
    \label{fig:RBM_eig_flow}
\end{figure}

The drift in Eq.\ (\ref{eq:xRBM}) can be integrated to yield the potential 
$V_i(x_i) = -\int^{x_i} dx'\, K_i(x') =  -x_i^2/2\kappa_i -x_i-\mu^2\log\left(\mu^2-x_i\right)$.
With the initial matrix elements of $W$ drawn from a normal distribution, the process of learning can then be summarised neatly as the stochastic evolution of the eigenvalues of $X$ from the Marchenko-Pastur distribution initially, as shown in Fig.~\ref{fig:RBM_eig_flow} (left) to the Coulomb gas (\ref{eq:Cgas}) with the potential $V_i(x_i)$ after training, as shown in Fig.~\ref{fig:RBM_eig_flow} (right). See Ref.~\cite{Aarts:2024wxi} for more details.

\subsection{Transformer}

In the Gaussian RBM the drift and potential in the Coulomb gas are known, resulting in full analytic control. In more involved architectures, this is typically not the case. Here we consider a nano-GPT (Generative Pretrained Transformer \cite{transformer}), with 4 attention blocks each having 4 attention heads. The total number of parameters is around 0.21M. Each attention head contains one key ($K$), one query ($Q$) and one value ($V$) matrix of shape $M\times N = 64\times 16$, where $M$ is the input feature dimension. 
The model we use is based on Karpathy's nano-GPT \cite{nanoGPT}, using the AdamW optimiser \cite{kingma:2017}. With highly adaptive stepsizes during training, the dependence on learning rate and batch size is more involved. Hence we explore here the evolution of the eigenvalue distribution of $X = W^TW$ through training, 
where $W$ is one of the attention matrices, at an empirical level.

The initial matrix elements are drawn from a uniform distribution, bounded between $\pm 1/\sqrt{M}$. The eigenvalue distribution of $X$ at initialisation is then given by an Marchenko-Pastur (MP) distribution 
\be
 \label{eq:MP}
 P_{\rm MP}(x; \sigma^2, A) = 
  \frac{A}{2 \pi \sigma^2 r x} \sqrt{(x_+ - x)(x - x_-)} \, \theta(x_+-x)\theta(x-x_-),
\ee     
with $x_\pm = \sigma^2 (1 \pm \sqrt{r})^2$,  $r=N/M=1/4$, area $A=1$ and $\sigma^2=1/3$ for the uniform initialisation.

\begin{figure}[t]
\begin{center}
   \includegraphics[width=0.32\textwidth]{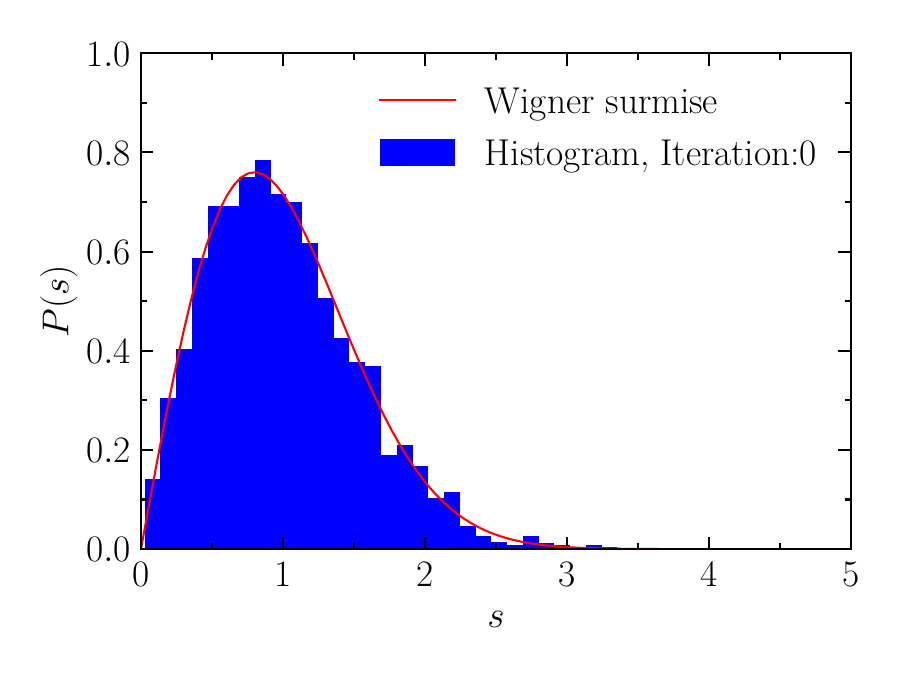}
    \includegraphics[width=0.32\textwidth]{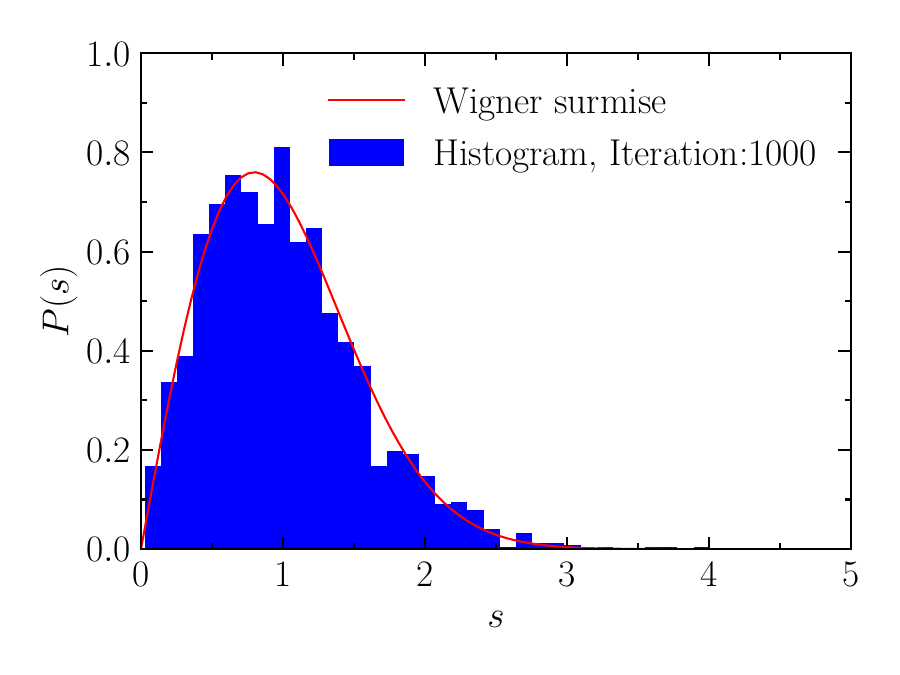}
    \includegraphics[width=0.32\textwidth]{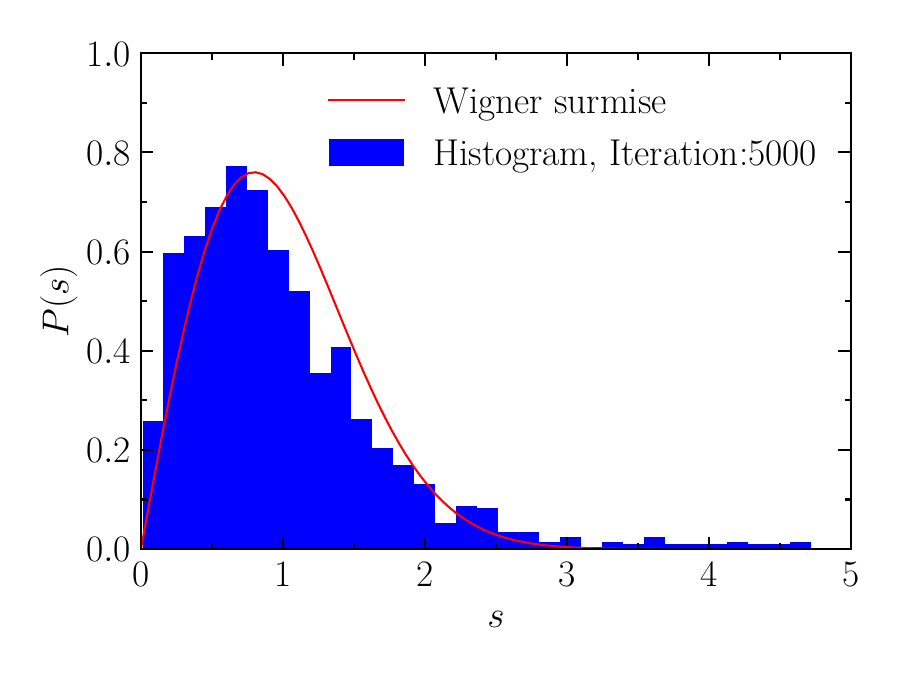}

    \includegraphics[width=0.32\textwidth]{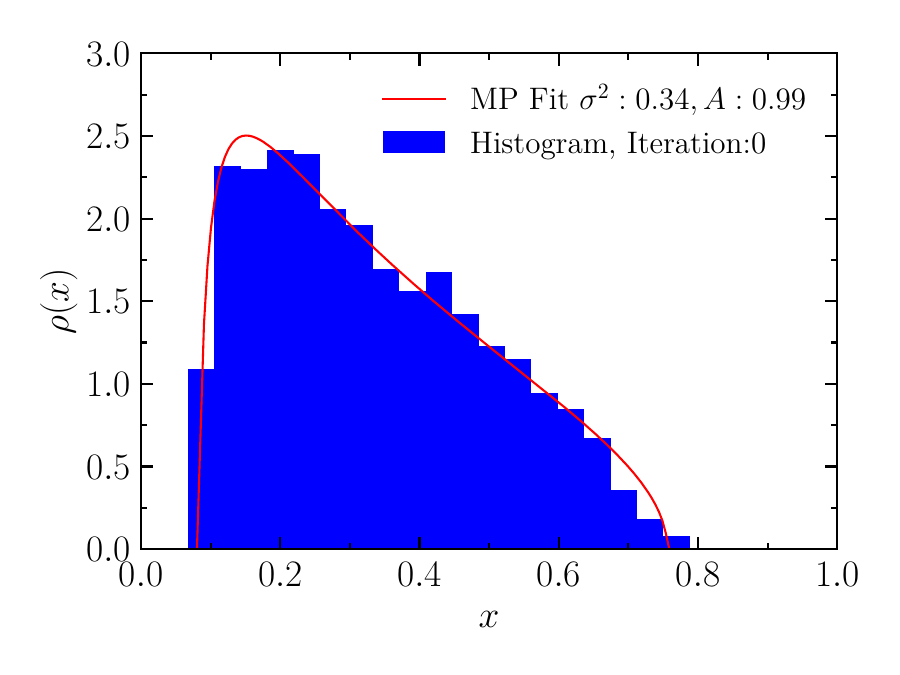}
    \includegraphics[width=0.32\textwidth]{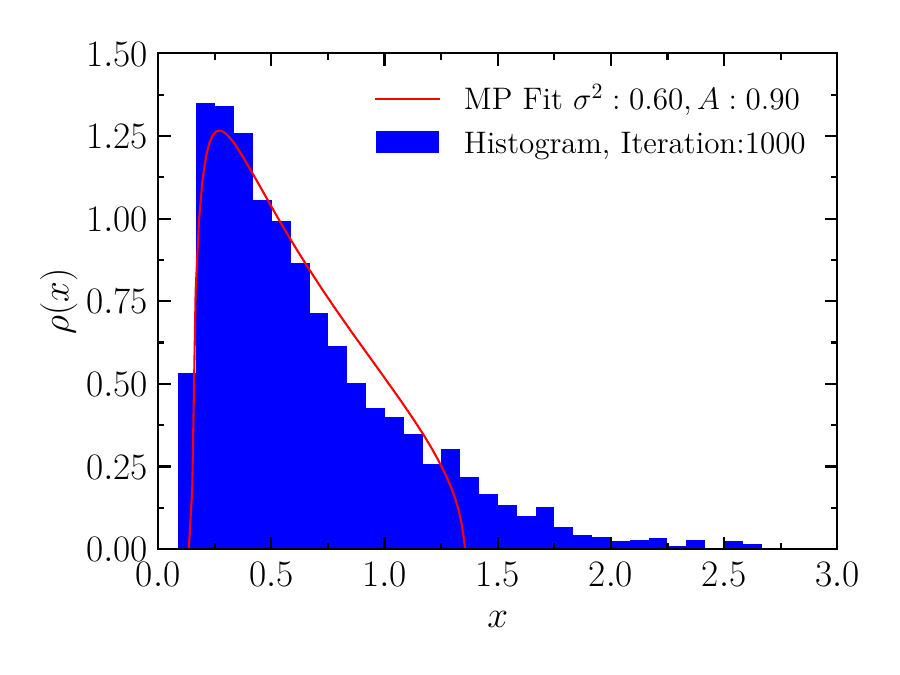}
    \includegraphics[width=0.32\textwidth]{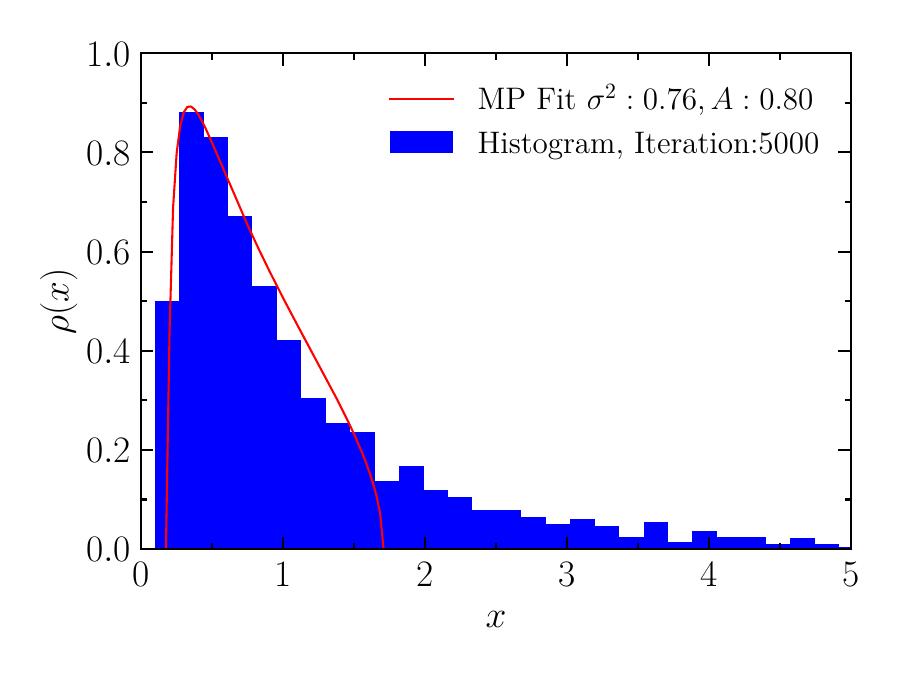}

    \caption{Transformer: Evolution during training of the eigenvalue distribution of $X=K^TK$, where $K$ is the Key matrix of the transformer's first layer, at initialisation (left), iteration 1000 (middle) and iteration 5000 (right). Above: distribution $P(s)$ of the normalised eigenvalue spacing $s_i=x_{i+1}-x_i$ after spectral unfolding, compared to the Wigner surmise. 
    Below: spectral density $\rho(x)$, compared to fits to the Marchenko-Pastur distribution with fit parameters $\sigma^2$ and area $A$. 
    }
    \label{fig:key}
\end{center}
\begin{center}
    \includegraphics[width=0.45\textwidth]{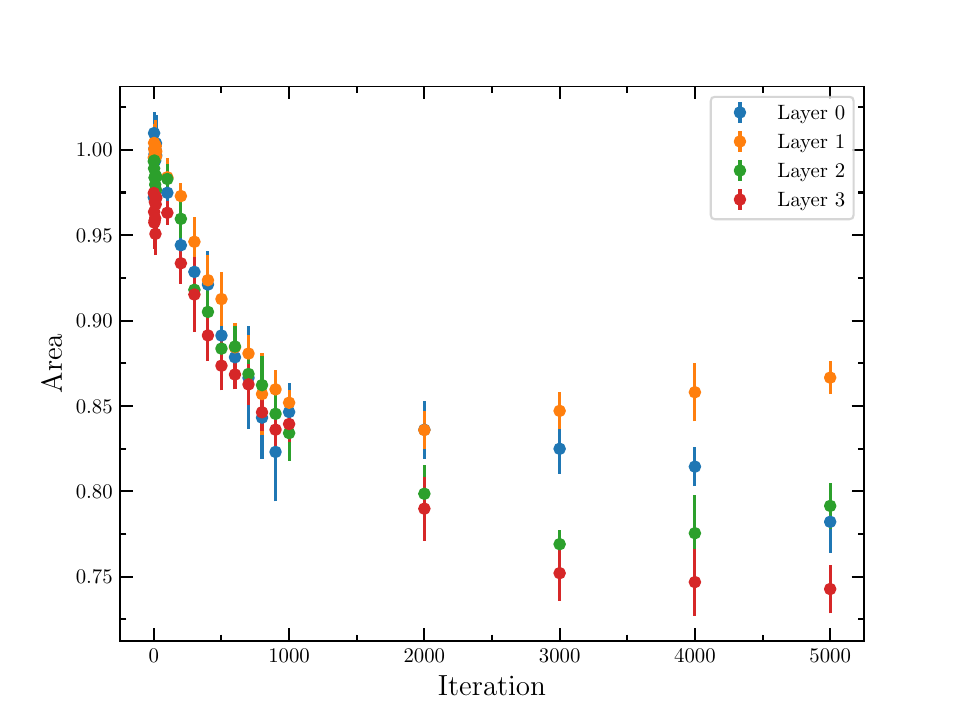}
    \includegraphics[width=0.45\textwidth]{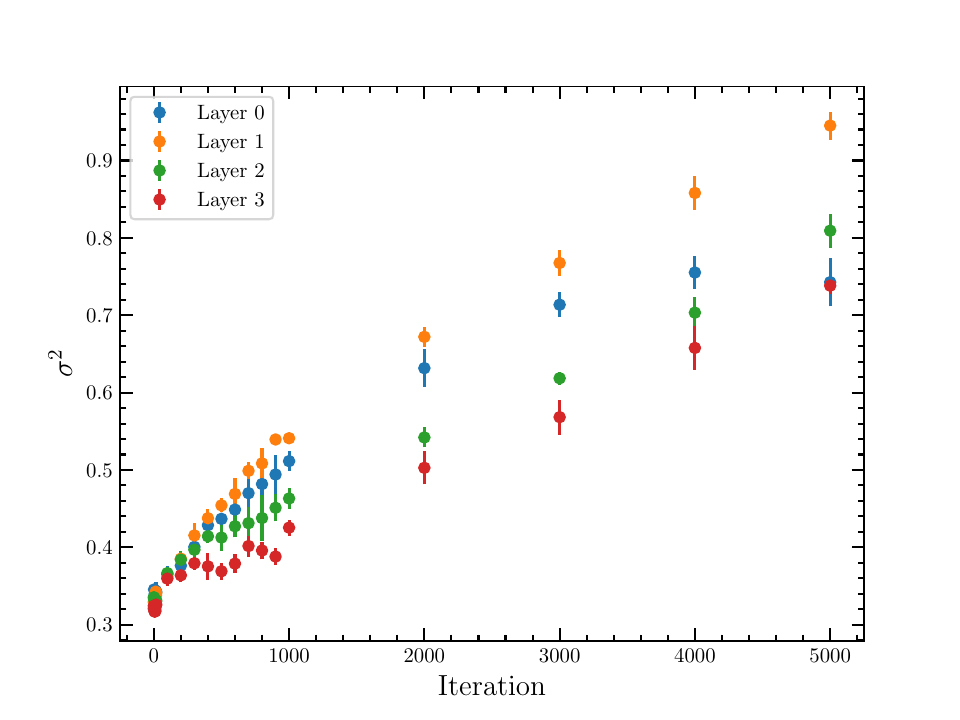}
    \caption{Transformer: Evolution of fit parameters area $A$ (left) and $\sigma^2$ (right) of the Marchenko-Pastur distribution fit to the spectral density $\rho(x)$ of $X=K^TK$ of the first head for all layers.
    To determine the statistical uncertainty, training is repeated at least 50 times, using a bootstrap analysis.
    }
    \label{fig:fit}
\end{center}
\end{figure}

In Fig.~\ref{fig:key} we show the evolution of the eigenvalue distribution of $X=K^TK$, where $K$ is the Key matrix of the transformer's first layer. 
To verify universal aspects, we show in the top row the distribution $P(s)$ of the normalised eigenvalue spacing, $s_i=x_{i+1}-x_i$, after spectral unfolding. The histograms are compared to the Wigner surmise, $P(s) = (\pi/2) s \exp(-\pi s^2/4)$, observing good agreement. This indicates that the eigenvalues undergo universal RMT fluctuations.

In the bottom row the evolution of the spectral density $\rho(x)$ is shown. From the initial MP distribution, one observes a shift of the spectrum to larger values, no longer described by an MP distribution (note the different horizontal scales). 
The appearance of (heavy) tails has been noted also in other architectures \cite{Martin-2019}.
This evolution should be compared to the evolution in the RBM in Fig.~\ref{fig:RBM_eig_flow}, with the notable difference that the ``exact'' spectrum or Coulomb gas potential are not known.
To quantify the evolution, we fit the spectral density to the MP distribution with fit parameters $\sigma^2$ and area $A$. The evolution of those is shown in Fig.~\ref{fig:fit} for the Key matrix of the first head for the four layers. 
As more spectral weight moves to the tail, the area $A$ of the part of the distribution still described by the MP distribution decreases. This is demonstrated in Fig.~\ref{fig:fit} (left), from which one deduces that the tail carries 15-25\% of the spectral weight.
The MP distribution itself evolves, as is shown in Fig.~\ref{fig:fit} (right), with the $\sigma^2$ parameter increasing from 1/3 to closer to 1, depending on the layer. This may reflect further randomisation of the lower part of the spectrum, with 
Brownian motion leading to an MP distribution with a larger domain (larger $x_+\sim \sigma^2$).
Work to further understand the spectral form of the tail is in progress, as is a study of the apparent non-monotonic behaviour of the area $A$.

\section{Summary}

We have argued that weight matrix dynamics should be viewed in the framework of Dyson Brownian motion. This predicts a universal dependence on the ratio of learning rate over batch size. It elegantly leads to a picture in which the process of learning is described as the stochastic evolution of the eigenvalues of $X=W^TW$ from an initial Marchenko-Pastur distribution to a Coulomb gas distribution with a model-dependent potential. This picture is fully confirmed in the Gaussian restricted Boltzmann machine and supported empirically in a nano-GPT. Further analysis of the latter is in progress.

\newpage

\noindent
{\bf Acknowledgements} --  
Part of this work was initiated at the ECT* workshop {\it Machine Learning and the Renormalisation Group} in May 2024. We thank the participants of this meeting for discussion and ECT* for support. 
GA and BL are supported by STFC Consolidated Grant ST/T000813/1. 
BL is further supported by the UKRI EPSRC ExCALIBUR ExaTEPP project EP/X017168/1.
CP is supported by the UKRI AIMLAC CDT EP/S023992/1.

\noindent
{\bf Research Data and Code Access} --
The code and data used for the first part of this manuscript are available from Ref.\ \cite{park_2024_13310439}.

\noindent
{\bf Open Access Statement} -- For the purpose of open access, the authors have applied a Creative Commons Attribution (CC BY) licence to any Author Accepted Manuscript version arising.

\providecommand{\href}[2]{#2}\begingroup\raggedright\endgroup

\end{document}